\documentclass[12pt]{JHEP3}

\title{Initial conditions for small-field inflation}

\preprint{}

\author{Micha\l\ Spali\'nski\footnote{Email: mspal@fuw.edu.pl}\\ 
So\l tan Institute for Nuclear Studies, Warsaw, Poland \\
and Physics Department, University of Bia\l ystok, Bia\l ystok, Poland.
}

\abstract{Small-field inflation typically requires rather special initial
  conditions to commence. It is proposed that in models where the
  inflaton is an axionlike field, with a periodic contribution to the
  potential, there is a possibility of significantly enhancing the chances of
  inflation without any fine-tuning of initial conditions and with no
  additional fine-tuning of the dynamics beyond what is needed for the potential
  to support inflation in the first place.  }

\keywords{Inflation}

\begin{document}

\newcommand{\spacer}{\vspace{6 pt}}
\newcommand{\bea}{\begin{eqnarray}}
\newcommand{\beal}[1]{\begin{eqnarray}\label{#1}}
\newcommand{\eea}{\end{eqnarray}} 
\newcommand{\be}{\begin{equation}} 
\newcommand{\bel}[1]{\begin{equation}\label{#1}}
\newcommand{\ee}{\end{equation}} 
\newcommand{\rf}[1]{(\ref{#1})}
\newcommand{\nn}{\nonumber}
\newcommand{\bit}{\begin{itemize}}
\newcommand{\eit}{\end{itemize}}
\newcommand{\ben}{\begin{enumerate}}
\newcommand{\een}{\end{enumerate}}
\newcommand{\no}{\noindent}
\newcommand{\req}[1]{(\ref{#1})}
\newcommand{\todo}[1]{\noindent {\bf TODO:} #1}
\providecommand{\eng}[2]{\makebox{\ensuremath{#1\times 10^{#2}}}}

\def\t{\tilde}
\def\del{\partial}
\def\d{\partial}
\def\half{\frac{1}{2}}
\def\third{\frac{1}{3}}
\def\quart{\frac{1}{4}}
\def\alp{\leavevmode\ifmmode {\alpha^\prime} \else ${\alpha^\prime}$ \fi}
\def\mps{M_P^2}

\section{Introduction}

It is often said that the enormous expansion of the Universe during
inflation erases
any memory of its pre-inflationary state. While this is
likely to be true, it is also the case that in many models of
inflation one needs rather special initial conditions for inflation to
start. This fine-tuning of initial conditions is quite separate from any
fine-tuning of the dynamics that may be necessary. 

Not all inflationary scenarios are afflicted with this malady; notably chaotic
inflation \cite{Linde:1983gd,Linde:1986fd} requires essentially no fine-tuning
of initial conditions. Models of 
this sort (frequently referred to as large-field models) are attractive
partly due to their simplicity. They are also of great
phenomenological interest, as they can easily accommodate increasingly accurate
observational data. Furthermore, since the inflaton traverses a trans-Planckian 
distance during inflation, large-field models typically predict potentially
measurable levels of tensor perturbations (as suggested by the Lyth bound
\cite{Lyth:1996im}). In some sense, the 
simplest models of chaotic inflation are the most attractive incarnation of the
idea of inflation. On the theoretical side
there are however some puzzles 
due to the necessarily trans-Planckian expectation values of the inflaton. Some
of the arguments appearing in this context invoke effective field theory ideas
while others reflect expectations based on string theory. 

There are also many models of inflation which do not involve trans-Planckian
expectation values of the inflaton (such models are sometimes referred to as
small-field models). An interesting class of inflationary models of the
small-field type has recently been discussed in the context of $D$-brane
inflation   
\cite{Baumann:2007np,Baumann:2007ah,Krause:2007jk,Cline:2009pu}, in the
supersymmetric standard
model \cite{Allahverdi:2006iq,Lyth:2006ec,Allahverdi:2008bt} and in supergravity 
\cite{Lalak:2007rsa,Badziak:2008gv}.  These scenarios envision inflation taking 
place close to an inflection point of the inflaton potential. In
each of these cases some fine-tuning of the effective inflaton potential is
involved; a recent treatment of these issues \cite{Linde:2007jn,Baumann:2007ah}
considers inflection points arising when the inflaton potential has a pair of
extrema which can be tuned to coincide. This requires fine-tuning at the level
of one part in $10^{-3}$.

If one accepts as inevitable the fine-tuning of the potential, inflection point
inflationary scenarios have a number of attractive features, such as the fact
that they admit a low scale of inflation which may be desirable from some points
of view. As a small-field model, inflection point inflation predicts a very low
level of tensor perturbations, which can be regarded either as a positive or
negative consequence. In the end this question will be settled by observation
\cite{Baumann:2008aq}.

A serious conceptual drawback of inflection point inflation is that it requires
fine-tuning of initial conditions as well as of the dynamics. An aspect of this
is often referred to as the overshoot problem \cite{Brustein:1992nk}: for
generic initial conditions the evolving inflaton field will simply miss the
inflection point as if it were not there at all 
\cite{Baumann:2007ah,Krause:2007jk,Itzhaki:2007nk}. The only way for inflation
to start is if some agent causes the inflaton expectation value to be very close
to the inflection point with negligible velocity. This is a potentially
disturbing aspect of this class of models\footnote{The problem of initial 
  conditions has recently been discussed in the context of hybrid inflation in
  \cite{Clesse:2008pf}.}.  It is the purpose of this short paper to suggest a
small-field scenario which does not involve any fine-tuning of the initial conditions,
and no additional fine-tuning of the dynamics beyond what is required to have an
approximate saddle point.

\section{Inflection point inflation}

Inflation near an inflection point has been discussed in a number of recent
articles \cite{Baumann:2007ah,Krause:2007jk,Linde:2007jn}, so only a very brief
summary is included here. Consider the scalar
potential $V(\phi)$ in the vicinity of a 
point $\phi_0$, where it 
is assumed that $V'(\phi_0)=V''(\phi_0)=0$. In this region the potential can be
approximated by
\bel{vip}
V(\phi) = V_0 + V_1 (\phi-\phi_0) +  V_3 (\phi-\phi_0)^3 + \dots
\ee
where $V_1=0$ for an exact saddle point, but the linear term could be
present without spoiling anything provided it is small enough. 
Allowing for $V_1\neq0$, to get at least 60 e-folds one needs 
\cite{Baumann:2007ah,Linde:2007jn}: 
\bel{finetune}
3 M_P^4 V_1 V_3 < 10^{-3} V_0^2
\ee
To simplify the formulae in this section $V_1$ will be neglected: more general
expressions can be found in \cite{Baumann:2007ah,Linde:2007jn}. 

The slow roll parameters are given by 
\bea
\epsilon &=& \half\mps (\frac{V'}{V})^2 \approx \frac{9}{2} \mps
(\frac{V_3}{V_0})^2 (\phi-\phi_0)^4\\
\eta &=& \half\mps \frac{V''}{V} \approx \mps\frac{6V_3}{V_0}(\phi-\phi_0)
\eea
Clearly one has $\epsilon\ll\eta$ in models of this type. 

In the slow roll approximation the number of e-folds is given
by 
\be
N = \frac{1}{\mps} \int_{\phi_i}^{\phi} d\phi  \frac{V}{V'} \approx
\frac{1}{\mps} \frac{V_0}{3V_3}\frac{1}{\phi_0-\phi}  \ ,
\ee
and the scalar perturbation spectrum is characterized by the spectral index  
\be
n_s-1 = 1 -6\epsilon +2\eta 
\approx 1-\frac{4}{N} \ .
\ee
Taking $N=60$ here and below (with standard assumptions about reheating) this
gives $n_s=0.93$, which is consistent with current limits. In the slow roll
approximation the 
amplitude of the scalar perturbation spectrum  is given by 
\be
P_S^2 = \frac{1}{12\pi^2M_P^6}\frac{V^3}{V'^2}
\ee
evaluated at horizon crossing. For the present case, using $P_S^2 = 2.5\times
10^{-9}$ \cite{Alabidi:2008ej}, this gives the condition  
\bel{cobe}
\frac{V_3^2}{V_0} = \frac{4\pi^2}{3 M_P^2 N^4} P_S^2 =  
\frac{2.5\times 10^{-15}}{M_P^2}  \ .
\ee

Another piece of information is provided by limits on the ratio of tensor to
scalar perturbations $r = 16 \epsilon$. Imposing $r < 0.25$ 
\cite{Alabidi:2008ej} translates into  
\bel{tenbound}
 \frac{V_0^2}{V_3^2} = \frac{9}{8} M_P^6 N^4 r < 3.6\times
  10^{6} M_P^6  \ .
\ee

The analysis summarized above (using the slow roll approximation) is somewhat
simplified -- to get a realistic picture one should also account for quantum
effects near the inflection point \cite{Krause:2007jk,Linde:2007jn}.

The major difficulty, as discussed in the introduction, is to motivate the
rather stringent initial conditions required for inflation to start.  Numerical
investigation shows that to avoid overshoot $\phi$ needs to be sufficiently
close to the inflection point, and furthermore $\dot{\phi}$ has to be
negligible.  In short, the initial condition needs to be in a suitable, small
region of phase space. It is clearly a problem to explain why the inflaton field
expectation value would reside in such a special region in the pre-inflationary
epoch. The scenario proposed in the following section is basically trying to
make this region of phase space appear less special.

\section{The ``staircase'' scenario}

The basic idea is that the dynamics of the inflaton is governed by a potential
which over some range of field space can be visualized as an ascending
staircase: an array of inflection points. A suitable potential could arise as
follows. Suppose that in some region of field
space the potential energy is dominated by two types of contributions: one of
them periodic and the other linear in the inflaton field. 
The periodic contribution is naturally interpreted as an instanton-generated 
energy density  of an axionlike field identified with the inflaton. This  
contribution, periodic with period $2\pi f$, defines a scale. The other key
element of the scenario 
is the presence of a second, nonperiodic\footnote{It could actually be
  periodic, as long as the period is much longer than $2\pi f$.} contribution 
$u(\phi)$, so that the energy density has the form  
\bel{edens}
V(\phi) = \mu^4 P\left(\frac{\phi}{f}\right) + u(\phi)  \ ,
\ee
where $P$ is a periodic function (with period $2\pi$) and $\mu$ is a parameter
with dimension of mass. The present proposal assumes that the  
scale of variation of $u(\phi)$ is much smaller than that set by $f$. If this is 
the case, it 
will be a good approximation to linearize this contribution over some
potentially large number of cycles of the inflaton. If there is enough freedom
in a specific model for these two contributions to be fine-tuned, the resulting
potential acquires, due to the periodicity of the first contribution, a sequence
of inflection points. The total
potential in this region resembles a smoothed staircase.   

An idealized realization, which can be envisaged to appear as an approximation
in many instances (including the specific examples described below) is to assume
the inflaton potential in the form 
\bel{proto}
V(\phi) = \mu^4 \left(- \sin \left(\frac{\phi}{f}\right) + (1+\alpha)
  \frac{\phi}{f}  + 
  \lambda\right) \ ,
\ee
where $\alpha$ and $\lambda$ are constant parameters. 
If it is possible to tune $\alpha \ll 1$, the potential has a
sequence of approximate saddle points at 
\be
\phi_k = 2 k \pi f \ ,
\ee
with $k$ assuming values in a range of integers such that the $\phi_k$ 
lie in the region of field space where \rf{proto} is valid. Close to any of
these points the potential takes the form \rf{vip} with
\be
V_0 = \mu^4 (\lambda + 2 k \pi)\ , \quad
V_1 = \alpha \frac{\mu^4}{f} \ , \quad
V_3 = \frac{\mu^4}{6f^3} \ .
\ee
The question of precisely how much fine-tuning of $\alpha$ is required will be
revisited at the end of this section. 

In the context of chaotic inflation it is usually assumed that in the
pre-inflationary universe the distribution of the inflaton field is essentially
random, apart from the assumption that the energy density remains sub-Planckian
so that field theory notions may be applicable \cite{Linde:1983gd}. In the
original proposal for chaotic inflation a spacial domain of extent $L \sim
H^{-1}\sim M_P^{-1}$ is considered where the inflaton field is homogeneous. The
chaotic inflation scenario assumes a large initial inflaton expectation value
and negligible velocity, but one can also consider the situation where the
initial inflaton velocity is large. This will lead to an ascending trajectory
where the inflaton stops at some point before it starts rolling back down.  If
the velocity of the inflaton is sufficiently large\footnote{Obviously one would
  like to keep the energy density is significantly below the Planck scale, so
  that Hubble friction does not dominate.} to scale at least one step of the
``staircase'' and as long as there is a large number of steps, any initial
condition of this sort will cause the inflaton to stop somewhere on the
staircase.  If the 
density of inflection points is sufficient, this turning point will be close
enough to one of them {\em with vanishing velocity} so that (at least
intuitively), there is a very good chance that inflation will commence. Making
that last statement precise would require adopting a satisfactory notion of
measure on the space of initial conditions.
  
If inflation takes place at the $k$-th step one can impose current observational
bounds to see if the constraints are reasonable. The condition on the tensor
ratio \rf{tenbound} gives
\bel{fbound} 
(\lambda + 2 k \pi)^2 \left(\frac{f}{M_P}\right)^6 <  1.0 \times 10^5
\ee 
which is a very weak bound: already for $k$
of the order 10 and $\lambda$ of order 1 this implies the axion decay constant
$f$ has to be below the Planck scale, which is in any case to be expected. Next,
applying the COBE normalization \rf{cobe} yields 
\bel{mubound} 
\mu^4 M_P^2 = f^6 (\lambda + 2 k \pi)\ 9.0 \times 10^{-14} \ .  
\ee 
This fixes the scale of the
potential $\mu$ well below the Planck scale for sensible values of the axion
decay constant and assuming that the parameter $\lambda$ is not large. This is a
reasonable assumption, given that large enough $\lambda$ would lead to a model
of chaotic inflation which would be analyzed differently. It is clear from
eq. \rf{proto} that $\mu$ determines the spacing of the steps 
in energy. Since the constraints \rf{fbound}, \rf{mubound} limit $\mu$ from
above, there is no obstruction from 
the observational side to
making the steps in energy quite dense.

One also needs to revisit the fine-tuning condition \rf{finetune}, which can be
expressed as 
\bel{alphatune}
\alpha < \third \left(\frac{f}{M_P}\right)^4 (\lambda + 2 k \pi)^2 10^{-3} \ . 
\ee
For a given $k$ this is a fine-tuning condition on alpha. However, if one
regards $\alpha$, $f$ and $\lambda$ as determined by the underlying theory, 
the relation \rf{alphatune} can be interpreted as a condition on $k$,
 the number of ``stairs'' the inflaton has to scale for 
sustained inflation to commence. This depends on the initial 
conditions, so \rf{alphatune} can be 
reinterpreted as a requirement that the initial inflaton ``velocity'' be large
enough. 

Despite Hubble friction the scale of inflation depends not only on the
potential, but also on the initial conditions. There is however no need to tune
them: all that is required for inflation to commence is that the turning point
is in a region of field space where the approximate form of the potential
\rf{proto} is 
valid.  Note also that if \rf{proto} is an approximate form of a symmetric
potential (valid in some range of $\phi$), then the sign of the initial inflaton
velocity is not relevant.

It is interesting to ask whether any traces of the ``uphill'' phase can be
detectable. This is very unlikely indeed, since if inflation takes place at an
inflection point typically very large numbers of e-folds ensue 
\cite{Linde:2007jn}, much exceeding the 55-70 e-folds of expansion since
visible perturbations were generated. Thus, generically there should be no
memory 
of any initial transient.

\section{Post-inflationary evolution}

After the Universe inflates at the uppermost inflection point reached, the
inflaton expectation value continues to move down. The potential gradient
accelerates the 
inflaton, but since there is another inflection point nearby it is important to
know whether another stage of inflation is possible as the inflaton approaches
that point. To answer this question one needs to resort to numerical analysis. 
The evolution of the inflaton is governed by Einstein's equations, which 
reduce to
\bea
\ddot{\phi} + 3 H \dot{\phi} + V_{,\phi}(\phi) &=& 0 \ , \nonumber \\
3 M_P^2 H^2  &=& \half \dot{\phi}^2 + V(\phi) \ .
\eea
It is convenient to use the number of e-folds, $N$, as the evolution parameter
\cite{Linde:2007jn}. Using $dN = H dt$ and denoting derivatives with respect to
$N$ by a 
prime, the resulting equations can be written in first-order form as 
\bea
\phi' &=& u \ , \nonumber \\
u' &=& -\frac{1}{H^2} \left(\frac{u V(\phi)}{M_P^2} + V_{,\phi}(\phi) \right) \
, \nonumber\\ 
H' &=& - \frac{1}{2 M_P^2} H u^2
\eea
with the Friedman constraint 
\be
\left(3 M_P^2 - \half u^2\right) H^2 = V(\phi) \ .
\ee
It is straightforward to integrate these equations numerically and consider
inflaton velocities at the point where inflation ends. 
For example, taking one of the inflection points considered by Linde and
Westphal \cite{Linde:2007jn} with 
$V_0 = \eng{2.7}{-23}$, $V_1 = \eng{7.29}{-32}$ and $V_3 = \eng{1.0}{-20}$ (in
Planck units) one finds 
that if the inflaton starts at rest with an initial value not exceeding
\eng{7}{-5} one gets hundreds of e-folds. The inflaton velocity $u$ at
the time 
when inflation ends depends very weakly on the initial condition and is of the
order $\eng{3}{-2}$. This is much too large for inflation to commence at the
next inflection point; at the parameter values cited, the initial inflaton
velocity cannot exceed $10^{-4}$ for inflation to start. 
Exploring various
choices of parameters leads to the  
conclusion that on the way down, once the inflaton makes it
past the inflection point where inflation takes place, it attains a velocity
which causes all the remaining inflection points to be 
overshot. Thus only one stage of inflation is possible in this scenario: at the
uppermost inflection point reached. 

While the background evolution is essentially unaffected by the presence of 
inflection points on the way down, they may leave observable traces in the
Cosmic Microwave Background,
since they induce small variations of the inflaton velocity. Eventually the
staircase region will end (assuming that the nonperiodic contribution $u(\phi)$
in \rf{edens} was bounded below). To formulate a realistic scenario the total
potential should have a minimum such that the Universe can reheat.  This part of
the story depends on the behavior of $u(\phi)$ in the region of field space
where the linear approximation is no longer valid and thus will differ from
model to model.

\section{Examples}

To construct specific models one needs an axion (to be identified with the
inflaton) and an instanton-generated potential with a reasonable scale and
periodicity. This potential restricts the shift symmetry of the axion to a
discrete subgroup. The other essential ingredient is a source of additional
symmetry breaking which eliminates the residual discrete shift symmetry. It is
natural to consider models of axion inflation in the framework of string theory
(e.g. \cite{Kallosh:2007cc,Grimm:2007hs}). In fact, string theory vacua provide
a setting where the key ingredients of the staircase scenario are easily found,
since they typically involve numerous fields with periodic potentials. These
fields are usually called axions, since it is expected that the QCD axion which
resolves the strong CP problem can be found among them. String theory axions
arise by dualizing second-rank antisymmetric fields, which arise in the process
of compactification in heterotic as well as type II vacua.

Before considering string theory, one can of course formulate 
suitable models purely in field theory\footnote{A model with an axion and a
  symmetry breaking linear potential was considered by Abbott
  \cite{Abbott:1984qf}, but in 
  that case the linear term was required to be small.}. 
One simple way that a staircase potential could appear is if there
were two periodic contributions to the axion potential with disparate periods:   
\bel{twocos}
V = \mu^4 \cos\left(\frac{\phi}{f}\right) + \mu'^4 \cos \left(\frac{\phi}{f'} +
\gamma\right) 
\ee
where $\gamma$ is a phase. 
If $f\ll f'$, then the contribution with the 
longer period could be approximated by a linear potential over many 
cycles of the first term. This way a potential of the form \rf{proto}
appears. 
Models like this were discussed by Freese, Liu and
Spolyar \cite{Freese:2005kt} (and recently in \cite{Ashoorioon:2008pj}). In that
work the relative contribution of the 
periodic and quasilinear pieces were chosen such that a sequence  
of minima ensued, leading to a realization of ``chain inflation''
\cite{Freese:2004vs}. Here the two terms in \rf{twocos} would have to be  
fine-tuned to give a sequence of inflection points leading to a
staircase potential in some region of field space. The staircase is finite in
this example, so that the initial conditions, while not fine-tuned, need to be
such that the inflaton turns back in the staircase region, i.e. the initial
inflaton velocity could not be too large. 

String theory offers good prospects for finding staircase potentials. 
An example which has the right ingredients appeared quite
recently \cite{McAllister:2008hb} in connection with efforts to construct a
string theory model of 
large-field inflation. 
The inflaton is identified with an axion 
arising from a 2-form field integrated over a 2-cycle $\Sigma_2$ in the usual 
way \cite{Svrcek:2006yi}. The potential receives an instanton contribution (due
to Euclidean $D$1-branes) which gives rise to a standard periodic term.
The authors of \cite{McAllister:2008hb} have however also identified a
nonperiodic contribution due to branes 
wrapping the 2-cycle $\Sigma_2$ . This term (basically the Dirac-Born-Infeld
action of the wrapped brane) is nonperiodic -- it 
contributes to the energy increasing without bound as a function of the axion
field. The ensuing potential is of the form \rf{edens}; specifically 
\be 
V(\phi) = \mu^4 \cos\left(\frac{\phi}{f}\right) + \mu'^3 \sqrt{v^2 + \phi^2} \ , 
\ee 
where $\mu$, $\mu'$ and $v$ depend on the geometry. 
For inflaton values much larger than $v^2$ the potential again takes the form 
\rf{proto}. 
In \cite{McAllister:2008hb} the
geometry was assumed such that the instanton contribution to the potential
was suppressed relative to the nonperiodic term. The relative
magnitude of the two terms is determined by the geometry, but there appears to
be no obstruction to fine-tuning the coefficients so that the two terms lead to  
an infinite staircase for $\phi\gg v$. It would certainly be interesting to
verify this and explore the observable consequences of this class of models.

\section{Conclusions}

Inflation at an inflection point is an attractive implementation of the
inflationary paradigm, which appears in a number of contexts.  A major concern
with such models is ensuring that inflation actually begins.  The scenario
described here addresses this issue. The basic observation is that if there is
any chance at all of inflation starting in a model with a single point of
inflection, that chance should increase if there is a whole sequence of such
points. A reasonably natural way such a structure could come about is by
``tilting'' a periodic potential due to the breaking of discrete shift
symmetry. Apart from the case of inflation at inflection points, this could also
be relevant for other instances of small-field inflation (e.g.  hilltop
inflation \cite{Boubekeur:2005zm}). This note has focused on presenting the
universal aspects of the idea.

Apart from making inflection point inflation somewhat more plausible, the
scenario proposed here suggests observational consequences which could
eventually support or refute it. The main source of information is, of course,
the Cosmic Microwave Background.  The uphill evolution of the inflaton could leave 
traces in the CMB in the form of large deviations from scale invariance
\cite{Krause:2007jk}. This would however only be observable if the visible 
perturbations were generated right at the start of inflation, that is, if
inflation lasted only for about 60 e-folds, which is very unlikely in the
scenario discussed above. Beyond this possibility, one could expect clear
signatures if multiple inflection points were traversed by the inflaton after
inflation. Even though the background evolution in the ``downhill phase'' is
hardly affected by the inflection points, the inflaton velocity would 
vary periodically as it crosses the steps on the way down which would be reflected in
the CMB as $k$-dependent oscillations in the spectrum of primordial density
perturbations. The impact of a ``feature'' in the potential on the perturbation
spectra has been the subject of numerous studies
(e.g. \cite{Covi:2006ci,Joy:2008qd}). The observable effects of a sequence of
steps induced by a duality cascade in the context of brane inflation were 
analyzed recently by Bean et al. \cite{Bean:2008na}. The observational imprint
of inflaton oscillations of the inflaton potential was also recently considered
in \cite{Pahud:2008ae,Jain:2008dw}. These studies have considered various forms
of the potential, but not quite the kind discussed here.  Clearly, it would be
very interesting to explore the signatures of a staircase of inflection
points.

Finally, it would be important to construct explicit, consistent models of
``staircase inflation'' in string theory, perhaps by making the second of the
two examples discussed above more precise. At this point it is not completely
clear that staircase inflation can be realized in string theory, but the
essential elements required are certainly present given the ubiquity of axions
and the rather generic mechanism for breaking the shift symmetry identified in
\cite{McAllister:2008hb}.  Although it is not obvious {\em a priori}, it appears
that  
these models have enough freedom to attain the requisite fine-tuning while
retaining control over the approximations made.  Since string theory has many
axion type fields, it is a very natural setting for staircase inflation.

\spacer
\spacer

\centerline{{\bf Acknowledgements}}

I would like to thank Shamit Kachru and Liam McAllister for helpful comments. 


\end{document}